\documentclass{article}
\usepackage{cite}
\usepackage{amsmath,amssymb,amsfonts}
\usepackage{algorithmic}
\usepackage{textcomp}
\usepackage{graphicx}
\usepackage{graphicx} 
\begin{document}

\title{Reply to “Structural Vulnerability in Y00 Protocols”}
\author{Geraldo A. Barbosa \\ \small{KeyBITS Encryption Technologies LLC, Reston - VA, USA - www.keybits.tech }\\
\small{E-mail: GeraldoABarbosa@keybits.tech}  }

\date{6 January 2025}
\newcommand{\be}{\begin{equation}}
\newcommand{\ee}{\end{equation}}
\newcommand{\bea}{\begin{eqnarray}}
\newcommand{\eea}{\end{eqnarray}}


\maketitle

\begin{abstract}
The paper arXiv:2412.07300v1 [quant-ph] 10 Dec 2024, entitled ``Structural Vulnerability in Y00 Protocols", by Kentaro Imafuku analyzes ``Secure Communication
Using Mesoscopic Coherent States", Phys. Rev. Lett., 90:227901, Jun 2003, by Geraldo A. Barbosa, Eric Corndorf, Prem Kumar, and Horace P. Yuen. 
Imafuku states that in his analyzes a structural vulnerability was revealed that enables the leakage of secret information from the measurement results in the Y00 protocol. Furthermore, it is also stated that an even simpler model, but based on the same original principles, can be more efficient than the original Y00 protocol.
This reply shows that: 1) the alleged vulnerability is due to a misinterpretation of one result by Imafuku and 2) that the simple model proposed by him is more vulnerable than the original Y00.
\end{abstract}

\section{Introduction}
In the Y00 protocol \cite{BCKY}, a standard symmetric key cipher, produced by a pseudo-random generator (a deterministic hardware or software), produces ciphertext from input coded data. The coded data are then transmitted by a quantum optical channel to introduce quantum uncertainty among the coding bases used, aimed at frustrating attacks. Ref. \cite{Imafuku} reminds the reader that in Japan, this idea produced positive government reports (https://www8.cao.go.jp/cstp/stmain/pdf\\/20230314thinktank/seikabutsu/
shir you3-2-13.pdf) leading to large-scale \\projects related to Y00 protocols:\\ https://www.nedo.go.jp/content/100979743.pdf. Ref. \cite{Imafuku} is one more effort to test Y00's resilience. 

Calculations in Ref. \cite{Imafuku} are usually correct, but some of his interpretations are incorrect. This paper discusses: 1) the impracticality of the proposed attack in \cite{Imafuku}, due to the need for exponential computational resources, and 2) shows that the proposed Toy model cannot achieve better efficiency than the original Y00 protocol. This is demonstrated by a detailed calculation of a generic encoding scheme, from which the proposed Toy model is a particular case. This conclusion is also reinforced by the so-called Helstrom's bound \cite{Helstrom}. 

\section{Exponential resources lead to impracticality}
Section ``4.2 Description via Maximum Likelihood Method" in \cite{Imafuku} 
formulates an attack based on the maximum likelihood estimation to obtain the measured sequence $\overrightarrow{z}_N=\{z_1,...z_N\}$ given that secret information $\hat{s}$ defines the measured outcomes. See Eq. (31) in \cite{Imafuku} and subsequent calculations for $p\left(\overrightarrow{z}_N| \hat{s}\right)$ and, in particular, see Eq.(34).

The combination of exhaustive search on the secret information space $S$ for each candidate $s$ scales with the entropy $\sigma(s)$. The maximum likelihood method tries many combinations of possible $s$ to find an absolute minimum. This demands an exponential computation resource that makes the specific attack impractical.  However, the author seemingly tries to minimize this crucial problem.

Ref. \cite{Imafuku} also refers to \cite{Barbosa} which also uses physical noise to conceal signals. But that paper is dedicated to key distribution and not encryption as Y00 \cite{BCKY}. Although these have different objectives, some principles are similar.  The Probability of Error $P^E_e$ by an attacker was calculated using an appropriate formalism for an optical channel, also applicable to Y00. See Eqs. (6) and (10) in \cite{Barbosa} for the theoretical results. Fig. 3 in the same paper shows numerical examples emphasizing that an eavesdropper cannot obtain the individual bits sent, regardless of the precision of the device used.
Collectively obtaining a succession of $k$ ``secrets" collectively does not eliminate the inherent difficulty presented by each sent bit.  Due to the independence of signals (coding bases, noise, and message), this leads to ${(P^E_e)}^k$ and to the exponential resources needed.

Section 4.1 in \cite{Imafuku} uses the set $ {\mathcal{ B}}^{+}$ to designate bases around the measured position. Y00 systems are designed to have a level of noise in the channel such that it covers a large number of bases around a measured position. In other words, the number of bases that can potentially carry a bit 0 or bit 1 around any measured position is approximately equal,  so that the probability of deciding for a 0 or 1 is $\sim 1/2$. This reasoning is valid for each sent bit and, therefore, for a number $k$ of successive bits this implies that the probability to obtain all bits is exponentially decreasing $\sim\left( 1/2 \right)^k$. This implies that an attack based on the maximum likelihood estimate of obtaining the measured sequence is already thwarted. In addition to this simple observation, maximum likelihood methods involve large searches over combinations of trial values, increasing the stress on computational resources.

It is interesting that in the ``Conclusions" section \cite{Imafuku}, the author briefly comments on the potential for this impracticality, but concludes that a simpler ``Toy model" with two levels presents better security than the Y00 protocol: ``\textit{there is no valid justification for adopting a weaker protocol. Since Y00 is demonstrably weaker than the Toy protocol}".

\section{The Toy model does not reflect reality}

The probability of error $P^E_e$ by an attacker derived in \cite{Barbosa} is quite general and can be adapted for smaller numbers of levels or a grouping of bases, including the two-level model (Toy model) proposed by \cite{Imafuku}. 

The security of Y00 depends on the amount of noise in the optical channel that is carried in a coherent mesoscopic state. 

For a better follow-up by the readers, a brief recall of the calculation of $P^E_e$ when using an optical channel is given below (see Section \ref{Pe calculation}). It can be applied both to all values $M$ and includes the Toy model. Also, the two-level protocol that Ref. \cite{Imafuku} proposes is a simple two state problem  to which an exact solution exists, the Helstrom's bound \cite{Helstrom}. This bound is briefly sketched in Section \ref{sHelstrom}. These will be used to formulate a conclusion about the proposed Toy model.

\subsection{$P^E_e$ calculation}
\label{Pe calculation}
The quantum channel used is an optical fiber carrying a coherent mode with wave function $|\alpha \rangle$. In the Fock state or number state, $|\alpha \rangle$ is written
\bea
\label{state}
|\alpha \rangle=e^{-|\alpha|^2/2}\sum_{n=0}^{\infty}\frac{\alpha^n}{\sqrt{n!}}|n \rangle\:.
\eea

Modulation of the light signal can be done using phase modulators. As the components of the electric field along $x$ and $y$ are orthogonal, one may write the coherent state $|\alpha \rangle$ decomposed along these two directions:
\bea
|\alpha \rangle \rightarrow |\mu \alpha \rangle_x \otimes |\nu \alpha \rangle_y\:.
\eea
In a simplified way, the effect of the phase modulator is to introduce a phase difference between the components $x$ and $y$. If one specifies photon annihilation operators for photons polarized along $x$ by $a$ and along $y$ by $b$, one may use the quantum $z$-angular momentum operator $J_z$ to produce a phase difference between the two components (See Appendix ``B. STOKES PARAMETERS OF A NOISY FIELD" in \cite{BarbosaGraaf}):
\bea
J_z=\frac{1}{2}(a^{+}a-b^{+}b)\:.
\eea
$a$ and $b$ are the annihilation operators for polarized photons along $x$ and $y$, while $^{+}$ represents the adjoint of the operators. $a$ and $a^+$ have a commutator $[a,a^+]=1$ and similarly $[b,b^+]=1$, but operators representing photons with distinct polarizations commute. 

A rotation operator $R(\phi)$ that introduces the phase difference $\phi$ can be written:
\bea
R(\phi)=e^{-i J_z \phi}
\eea
and represents the action of the modulator on the incoming field. The out-coming state $|\Psi \rangle$ can be written ($\otimes$ will be omitted):
\bea
|\Psi \rangle=e^{-i J_z \phi} |\mu \alpha \rangle_x |\nu \alpha \rangle_y \:.
\eea
One gets
\bea
\label{Psi}
|\Psi \rangle=|\mu \alpha e^{-i\phi(j,k)/2}\rangle_x |\nu \alpha e^{i\phi(j,k)/2} \rangle_y \
\eea
Just for simplicity, one adopt $ \mu=\nu=1/\sqrt{2}$. The state $|\Psi \rangle$ describes the physical signal sent from A to B. The associated density matrix $\rho$ is
\bea
\rho=|\Psi \rangle  \langle \Psi|\:.
\eea

\subsection{Mixed state}
The attacker does not know which basis (or secret information $s$, as used by \cite{Imafuku}) was used at each instant $t$. Therefore,
the \underline{a-priori} {\em maximum} information at $t$ available for the attacker will be a {\em copy} of the original irreducible {\em noisy} signal, described by the density matrix of the mixed state
\bea
\label{rhotilde}
\tilde{\rho}_j=\frac{1}{M}\sum_{k=0}^{M-1} \rho_{j,k}\:\:.
\eea
The $\sum_{k=0}^{M-1}$ represents the bases (the set of angles $\phi_k \:\:(k=0,1,...M-1)$) information that are  hidden from the attacker and
\bea
\rho_{j,k}&=&|\Psi(j,k) \rangle \langle \Psi(j,k) |\nonumber \\
&=&|  \frac{ \alpha}{\sqrt{2}} e^{-i\phi(j,k)/2}  \rangle_x | \frac{ \alpha}{\sqrt{2}} e^{i\phi(j,k)/2} \rangle_y \nonumber \\
&&\otimes
\langle \frac{ \alpha}{\sqrt{2}} e^{-i\phi(j,k)/2} |_x \langle \frac{ \alpha}{\sqrt{2}} e^{i\phi(j,k)/2}|_y \nonumber \\
&=& e^{-iJ_z \phi_k}|  \frac{ \alpha}{\sqrt{2}}e^{-i\phi_j/2}   \rangle_x | \frac{ \alpha}{\sqrt{2}} e^{i\phi_j/2} \rangle_y \nonumber \\
&&\otimes
\langle \frac{ \alpha}{\sqrt{2}}e^{-i\phi_j/2} |_x \langle \frac{ \alpha}{\sqrt{2}}e^{i\phi_j/2} |_y e^{iJ_z \phi_k}\:
\:.
\eea

\subsection{Error for the attacker, E}

This section derives the case of an arbitrary $M$ that, at the end, can be simplified for $M=2$ (the Toy model in \cite{Imafuku}), for a general comparison.

As Fig. 4 in \cite{Imafuku} shows, each basis $b_j$ can carry either a transmitted signal 1 or 0 (signals describing bits).
 The sets are such that adjacent signals (or bits) are opposed. That is to say, the bits at closest range of a given bit are opposite bits. In a phase modulation, bits 0 and 1 are separated, for each basis, by an angle $\pi$.

One may ask what would be the best result that the attacker's measurement apparatus could achieve in the $M$-ry situation or, alternatively, what the minimum probability of error is for this case. Even considering perfect performance of the instrumentation, the noisy channel produces uncertainties that are irreducible by principle and affect the identification of the basis being used.

  In general, even without a description of the instrumentation that the attacker could use, one may assume that she is able to make projections on the even or odd bases using projectors $\Pi_0$ and $\Pi_1$. These projectors form a complete set that covers all signal possibilities that could be generated by the system. Therefore, $\Pi_0+\Pi_1=I$.

The failure of the correct basis to get a bit, taking into account all possible bases, can be represented by the operator
\bea
\mathcal{P_M}=\Pi_0 \rho_1+\Pi_1 \rho_0\:.
\eea

The operator $P_M$, describes positions 0 in the bases when the bit sent was bit 1 and vice versa.

Therefore, the minimum probability or error for E, considering that bits are chosen randomly with probability $p_0=p_1=1/2$, is
\bea
\label{min}
P_e^E&=&\frac{1}{2}Tr \mathcal{P_M}=\frac{1}{2} Tr\left[ \Pi_0 \tilde{\rho_1} +\Pi_1 \tilde{\rho_0}   \right]\nonumber \\
&=&
\frac{1}{2}Tr\left[(I-\Pi_1) \tilde{\rho_1} +\Pi_0 \tilde{\rho_1}   \right]\nonumber \\
&=&
\frac{1}{2}Tr\left[ \tilde{\rho_1} -\Pi_1 \left(\tilde{\rho_1}-\tilde{\rho_0}\right)   \right]\nonumber \\
&=&
\frac{1}{2}\left(  Tr\left[ \tilde{\rho_1}\right] -Tr\left[ \Pi_1 \left(\Delta \rho \right)   \right] \right)\nonumber \\
&=&
\frac{1}{2}\left(  1 -Tr\left[ \Pi_1 \left(\Delta \tilde{\rho} \right)   \right] \right) \:,
\eea
where $\Delta \tilde{\rho}=\tilde{\rho_1}-\tilde{\rho_0}$ and $Tr\left[ \tilde{\rho_1}\right]=1$ .

The development of $\Delta \tilde{\rho}$ is straightforward but cumbersome. This is shown in the Appendix. The result obtained is that

\bea
\label{DeltaFinal}
\Delta \tilde{\rho}\!\!&=\!\!&\!\!\sum_{p=-\infty}^{\infty}
\sum_{q=-\infty}^{\infty} 2
 e^{- 2|\beta|^2}\left[\frac{1-(-1)^{q-p}}{2}  \right] \sqrt{I_{2p}\left(2 |\beta|^2  \right) I_{2q}\left(2 |\beta|^2  \right)}  \times S(p,q) \times \nonumber \\
&& |\Phi_p\rangle \rangle \langle \langle\Phi_p |\:,
\eea
where $I_j$ is a modfified Bessel function.
\bea
\label{Spq}
S(p,q)\equiv \sum_{k=0}^{M-1}e^{-i \phi_k (q-p)}\:.
\eea

The analytical eigenvalues of $\Delta \tilde{\rho}_{p,q}$ would provide a complete
Probability of Error for the adversary (see Eq. (\ref{min})),  and the
eigenfunctions would reveal how the optimum measurement
could be achieved. 
$
\Delta \tilde{\rho}_{p,q}
$
are elements of a Hermitian matrix and, therefore,
with real eigenvalues, from which Eq. (\ref{min}) can be numerically calculated.

\section{Encryption bases}
\subsection{Full wheel configuration and probability of error}

The description of encryption bases in the phase modulation case can be calculated by
\bea
\label{angle}
\phi_k=\pi  \left[ \frac{k}{M}+\frac{1-(-1)^k}{2} \right]\:,
\eea
 where $k$ is one among $M$ bases.

Use of Eq.~(\ref{angle}) with a a full circle of phases (``full wheel"), from 0 to $2 \pi$, leads to diagrams such as Fig. (4) in \cite{Imafuku}.

Fig. 4 in \cite{Imafuku} shows encoding diagrams with bits $\{0,1\}$ and bases $b_j$. It is easy to see that in the phase encoding choice,  when states differ by a large angle, their identification is easier. Whenever the angle separation between states is small ($M$ large), the associated light noise frustrates state identification. This is calculated using Eq.~(\ref{min}), with the analytical result given by Eq. (\ref{DeltaFinal}).

\begin{figure}[h!]
\centerline{\scalebox{0.6}{\includegraphics{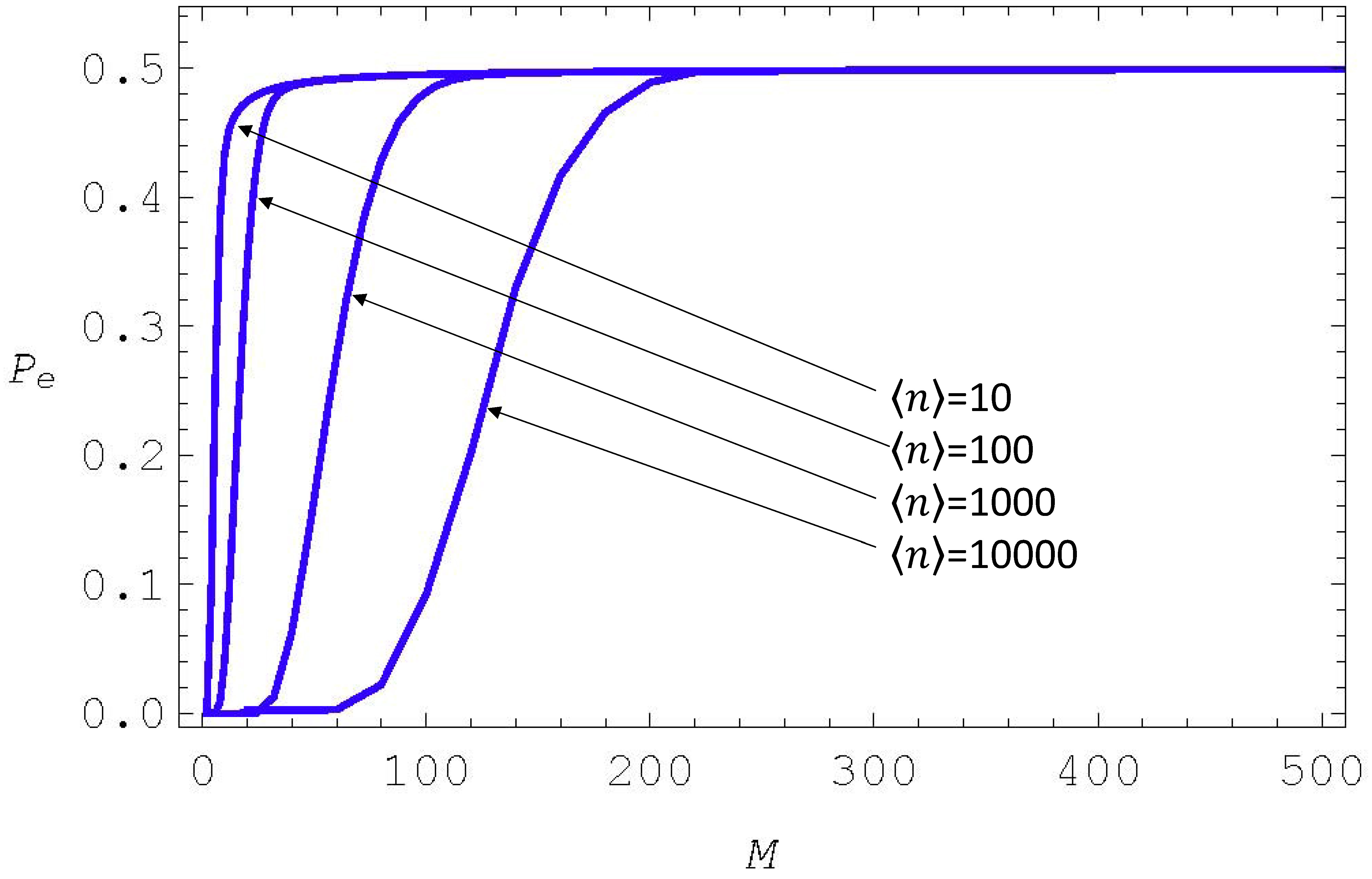}}}
\caption{ \label{PeFullWheel} Probability of Error for an attacker as as function of the number of bases used, $M$, and the average number of photons $\langle n \rangle$ carrying a bit. It is seen that as $M$ increases, $P_e \rightarrow 1/2$.}
\end{figure}

Fig.~\ref{PeFullWheel} gives a numerical example of the Probability of Error obtained from Eq. (\ref{DeltaFinal}), for the attacker who does not know which basis was used.

\subsection{Fan configuration and probability of error}
\begin{figure}[h!]
\centerline{\scalebox{0.4}{\includegraphics{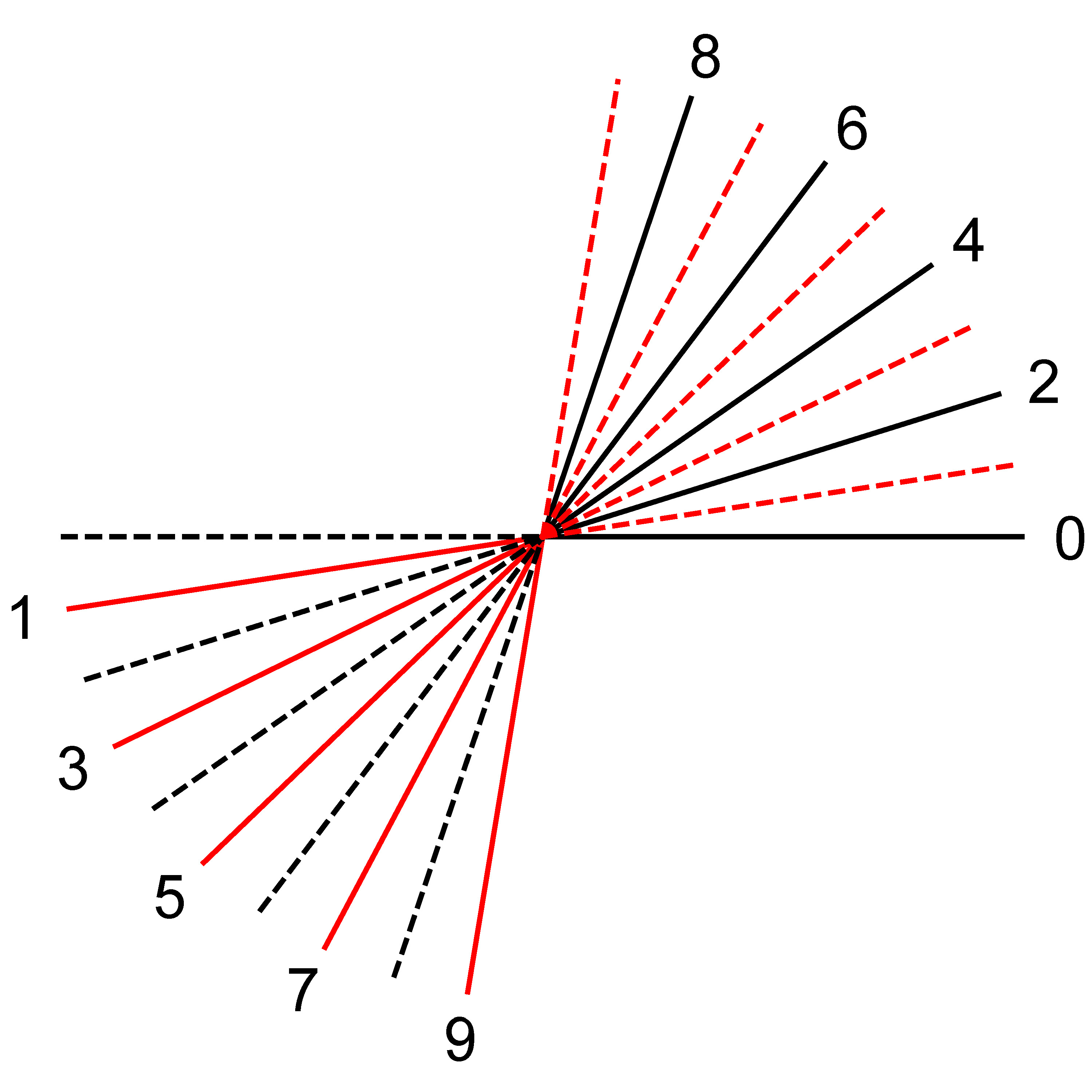}}}
\caption{ \label{Fan10} Fan of bases with $M=10$ and $f=2$. }
\end{figure}

For a spoke-spherical fan, Eq. (\ref{fan}) is written as a modification of Eq. (\ref{angle}):
\begin{eqnarray}
\label{fan}
\phi_{fk}=
 \pi   \left[\frac{k}{f M_f} +
\frac{1-(-1)^k}{2}  \right]\:, \:k=0,1, ...,M-1\:,
\end{eqnarray}
and defines a fan of spokes that does not encompass the whole circle. Rather, it is concentrated on $\pi/f$ of the circle (the full wheel is a particular case of Eq.~(\ref{fan}) with $f=1$). See Fig.~\ref{Fan10} for an example with $M=10$ and $f=2$.

 This shows that a spoke fan can be studied where the Toy model of \cite{Imafuku} is the simplest case.

As the light noise itself is independent of our choice of the encryption scheme, one may assume that a good guidance for choosing a fan of bases giving similar protection as a full wheel will be to select $f M_f\equiv N_B$, or $M_f=N_B/f$ as one fixes the span of a phase sector.

\begin{figure}[h!]
\centerline{\scalebox{0.5}{\includegraphics{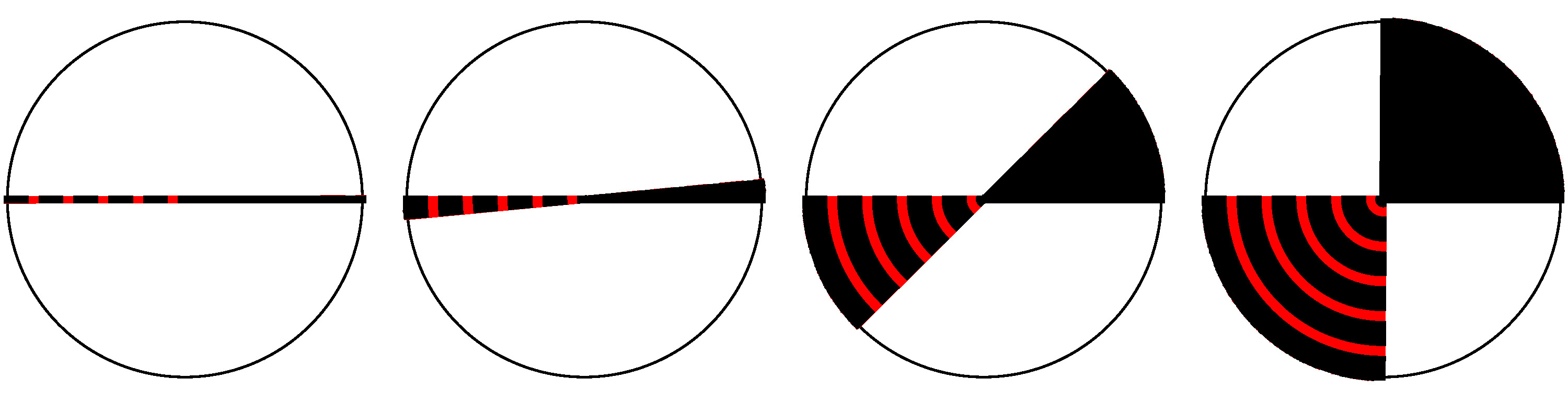}}}
\caption{ \label{gr} Set of fans with closest bases separation of $\pi/N_B$, with $N_{B}=2^{10}$ for fans with 2, 32, 256 and 512 bases.  }
\end{figure}

Fig.~\ref{gr} illustrates fans with 2, 32, 256 and 512 bases.

\subsection{Probability of error for fan configurations}

The calculation of $P_e$ differs slightly from Eq.~(\ref{DeltaFinal}) due to the term $S(p,q)$.  $e^{-i \pi(q-p)/M}$ should be written 
\bea
e^{-i \pi(q-p)/M} \rightarrow e^{-i \pi (q-p)/N_B}\:,
\eea
and $\phi_k$ would be written with $k=0,1,2,3,\dots M_f$, where $M_f$ is the number of bases in a given fan. $S(p,q)$ results into
\bea
&&S(p,q)=\frac{1}{2 M}\times\nonumber \\ &&
\left[ \frac{\left(1-(-y)^M\right)
   }{1+y}\!\left(1-y^{N_B}\right)
   +\frac{\left(1-y^M\right)
   }{1-y}\!\left(1+y^{N_B}\right) \right]\:. \nonumber
\eea
$M$ is a variable describing an arbitrary number of bases and $M_f$ as the maximum number of bases of the fan.

\begin{figure}[h!]
\centerline{\scalebox{0.5}{\includegraphics{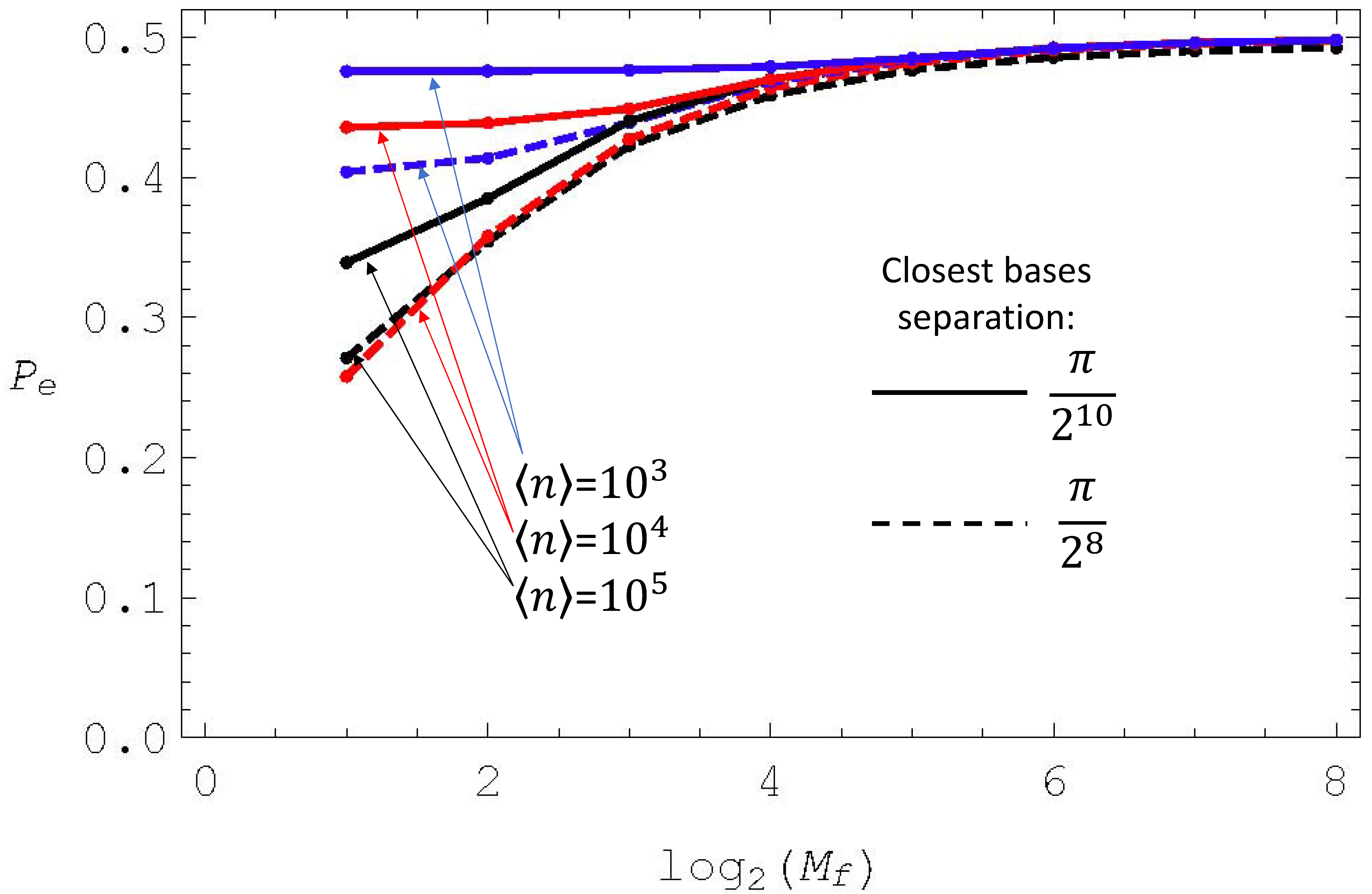}}}
\caption{ \label{AllFan} $P_e$ for set of fans with varying number of bases, varying separation between closest bases and for different average number of photons $\langle n \rangle$. Observe that for two levels ($M=2$),  $P_e$ is lower  compared with other $M$ values. This shows that the Toy model offers \underline{less} protection than Y00.}
\end{figure}
Fig.~\ref{AllFan} shows $P_e$ for sets of fans where in the first set (solid lines) the base separation is $\pi/N_B=\pi/2^{10}=\pi/1024$ rd and in the second set (dashed lines) the separation is $\pi/N_B=\pi/2^{8}=\pi/256$ rd. The upper solid curve (blue) is the set of data when the average number of photons carrying the bit is $\langle n \rangle=1000$. For the next solid curve (red), $\langle n \rangle=10000$ and for the black solid curve $\langle n \rangle=100000$.

Similarly, the upper dashed  curve (blue) is the set of data when the average number of photons carrying the bit is $\langle n \rangle=1000$. For the next dashed curve (red), $\langle n \rangle=10000$ and for the black dashed curve $\langle n \rangle=100000$.

It is clear that by increasing the number of levels $M_f$ in a fan, $P_e$ increases while, by increasing $\langle n \rangle$, $P_e$ decreases (intense light allows for a better resolution). As the level separation decreases, $P_e$ increases, which shows the effect of phase noise affecting the resolution of phases in a measurement.

For the calculation of the eigenvalues of $\Delta \tilde{\rho}(p,q)$ leading to $P_e$, square matrices with $(101)^2$ to $(2001)^2$ terms (values of $p$ and $q$) were used, where the highest number of terms was used for the largest separation of bases and the highest average number of photons.

Similarly as in the case of a full coding wheel, one should use a reasonably large number of bases in a fan together with a small separation of closest bases and with a reasonably small average number of photons, as formerly discussed.

These detailed results demonstrate that the Toy model given by \cite{Imafuku} is \textbf{not} superior to the original Y00 protocol with a uniform wheel of bases.
Basically, this \underline{invalidates the main conclusion of \cite{Imafuku}}.

\subsection{Helstrom's bound}
\label{sHelstrom}

\begin{figure}[h!]
\centerline{\scalebox{0.6}{\includegraphics{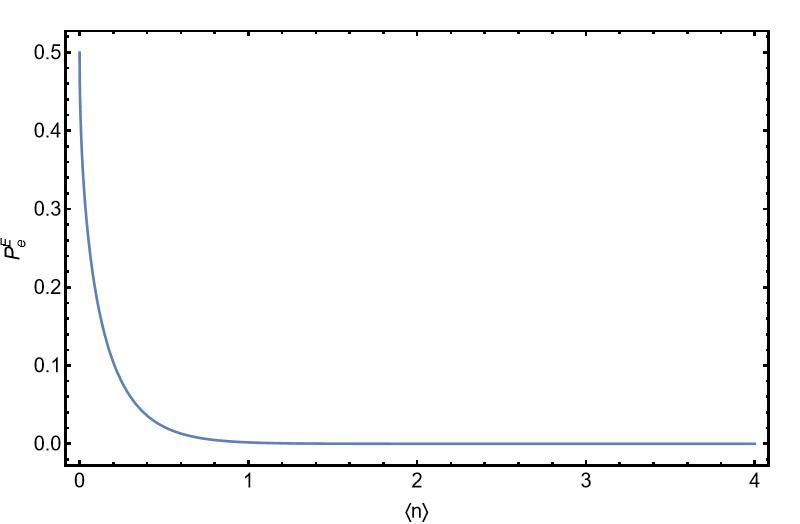}}}
\caption{ \label{pPerrH} ${P_e}^E$ for an attacker on transmission with two-level systems. $\langle n \rangle$ is the average number of photons in the carrier light beam.}
\end{figure}

The two-state discrimination problem has a classical analysis by Helstrom \cite{Helstrom}.  It shows that the Probability of Error ${P_e}^E$ by an attacker is:
\bea
{P_e}^E=\frac{1}{2}\left( 1-\sqrt{1-p_0 p_1 |\langle \Psi_0|\Psi_1\rangle|^2} \right)    \:,
\eea
where $\Psi_0$ and $\Psi_1$ are the two states and $p_0$ and $p_1$ their respective probabilities (here considered $p_0=p_1=1/2$). $|\langle \Psi_0|\Psi_1\rangle|^2$ is calculated in Ref. \cite{Barbosa} (Eq. (2)):
\bea
|\langle \Psi_0|\Psi_1\rangle|^2=e^{-2 \langle n \rangle   \left[ 1-\cos\left( \frac{\Phi_p -\Phi_k}{2}\right) \right] }
\simeq e^{- \langle n \rangle \frac{\Delta \Phi}{2}}\:.
\eea
This is the estimate of the uncertainty $\Delta \Phi$ of the polarization angle caused by the noise from the light beam. Fig. \ref{pPerrH} exemplifies ${P_e}^E$ as a function of the average number of photons in the optical channel.

Clearly ${P_e}^E$ goes to zero even with a small number of photons. This reinforces that using the two-state model is not an effective prescription if one wants to 
\textbf{protect} the information from the attacker. Some protection is provided only at $\langle n \rangle 
\leq 1$, but these low levels are not adequate for long-haul transmission due to light absorption in the channel. Quantum states cannot be cloned, and amplifiers are used for practical systems.
This reinforces the lack of practicality of the two-state model proposed in \cite{Imafuku}.

\section{Conclusions and beyond conclusions}

Ref. \cite{Imafuku} added to several studies of Y00, as given by the references cited. As the author emphasizes, encryption systems that may reach large-scale applications have to be studied in an exhaustive way. The author's conclusion that his simple Toy model has superior security than Y00 has been demonstrated to be wrong.

Certainly, new studies are welcomed and many questions have to be answered~\!\footnote{This reply is not including observations from the other authors of \cite{Barbosa}, which may also present their own points of view.}. One of these relates to the use of pseudo-random number generators for encryption as used in Y00. Superior computation capabilities -even including future quantum computation - suggest that the knowledge of inner workings (the formation rules for the random numbers in a pseudo-random number generator) may be fragilized by programs yet to be written.

The use of nondeterministic random generators is expected to increase for quantum-resistant encryption as well as for secure key distribution systems, as shown in \cite{KeyBITS}.

\section{Appendix}
\subsection{Developing $\Delta \tilde{\rho}$}

The mixed state given by Eq.~(\ref{rhotilde}) is composed of the density matrices representing the possible realizations of the bit encoding that could be performed by the emitter A:
\bea
\rho_{j,k}&=&|\Psi(j,k) \rangle \langle \Psi(j,k) |\nonumber \\
&=& e^{-iJ_z \phi_k}|  \frac{ \alpha}{\sqrt{2}}e^{-i\phi_j/2}   \rangle_x | \frac{ \alpha}{\sqrt{2}} e^{i\phi_j/2} \rangle_y \nonumber \\
&&\otimes
\langle \frac{ \alpha}{\sqrt{2}}e^{-i\phi_j/2} |_x \langle \frac{ \alpha}{\sqrt{2}}e^{i\phi_j/2} |_y e^{iJ_z \phi_k}\:
\:.
\eea
At each basis the encoding of bit 0 or 1 will be represented by the wave function
\bea
|\frac{ \alpha}{\sqrt{2}}e^{-i\phi_j/2}   \rangle_x | \frac{ \alpha}{\sqrt{2}} e^{i\phi_j/2} \rangle_y \:\:,\:(j=0,1)\:.
\eea
One may start developing the fundamental operator difference in Eq.~(\ref{min}), with the encoding of bits 1 and 0 (For short, $\beta$ replaces $\alpha/\sqrt{2}$):
\bea
\label{rho1rho0}
&&|  \beta e^{-i \pi/2}   \rangle_x |  \beta e^{i \pi/2} \rangle_y \otimes
\langle  \beta e^{-i \pi/2} |_x \langle  \beta e^{i \pi/2} |_y  \nonumber \\
&&-|  \beta e^{-i 0/2}   \rangle_x |  \beta e^{i 0/2} \rangle_y \otimes
\langle  \beta e^{-i 0/2} |_x \langle  \beta e^{i 0/2} |_y \nonumber \\
=&&| -i \beta    \rangle_x | i \beta  \rangle_y \otimes
\langle -i \beta  |_x \langle i \beta  |_y  \nonumber \\
&&-|  \beta    \rangle_x |  \beta \rangle_y \otimes
\langle  \beta  |_x \langle  \beta  |_y\:.
\eea
With the above one could look at the base encoding in Eq. (~\ref{rho1rho0}) and use  Eq.~(\ref{state}) to expand (\ref{rho1rho0}) in the number of bases or Fock states. The base encoding is
\bea
&&e^{-iJ_z \phi_k} \left[ | -i \beta    \rangle_x | i \beta  \rangle_y \otimes
\langle -i \beta  |_x \langle i \beta  |_y \right. \nonumber \\
&&\left. -|  \beta    \rangle_x |  \beta \rangle_y \otimes
\langle  \beta  |_x \langle  \beta  |_y  \right]  e^{iJ_z \phi_k}
\eea
Observe that
\bea
&&e^{-iJ_z \phi_k}  | -i \beta    \rangle_x | i \beta  \rangle_y=
| -i \beta e^{-i\phi_k/2}   \rangle_x | i \beta e^{i\phi_k/2} \rangle_y  \nonumber \\&&
=e^{-|-i \beta e^{-i\phi_k/2}|^2/2}\sum_n \frac{(-i \beta e^{-i\phi_k/2})^n}{\sqrt{n!}}|n\rangle \nonumber \\
&&\times
e^{-|i \beta e^{i\phi_k/2}|^2/2}\sum_m \frac{(i \beta e^{i\phi_k/2})^m}{\sqrt{m!}}|m\rangle\nonumber \\
&&=e^{-|\beta|^2}\!\!\sum_n\sum_m \frac{(-i\beta)^{n}(i \beta)^{m}e^{-i \phi_k(n-m)/2}}{\sqrt{n! m!}}|n\rangle |m \rangle\:.
\eea
Developing similar terms, the base encoding gives
\bea
\label{eqn}
&&e^{-iJ_z \phi_k} \left[ | -i \beta    \rangle_x | i \beta  \rangle_y \otimes
\langle -i \beta  |_x \langle i \beta  |_y \right. \nonumber \\
&&\left. -|  \beta    \rangle_x |  \beta \rangle_y \otimes
\langle  \beta  |_x \langle  \beta  |_y  \right]  e^{iJ_z \phi_k}\nonumber\\
&&=
2 i e^{-2|\beta|^2}\nonumber \\ &&\times\sum_{n_1}\sum_{n_2}\sum_{m_1}\sum_{m_2}
\frac{\beta^{n_1+n_2}(\beta^*)^{m_1+m_2}}{\sqrt{n_1!n_2!m_1!m_2!}}\nonumber \\
&& \times (-i)^{\frac{n_1+m_2}{2}}i^{\frac{n_2+m_1}{2}}  \sin\left( \frac{\pi}{4}
\left[ (n_2+m_1)-(n_1+m_2) \right]\right)
\nonumber \\&&\times e^{i\phi_k\left[(n_2-n_1)-(m_2-m_1)\right]}| n_1 \rangle  | n_2 \rangle \langle m_2| \langle m_1|\:.
\eea
Looking for possible simplifications, one may define 4 variables $J$, $J'$, $p$ and $q$ such that
\bea
&&n_1=J-p\:\:\:,\:\:\:m_1=J'-q \nonumber \\
&&n_2=J+p\:\:\:,\:\:\:m_2=J'+q \:,
\eea
and obtains
\bea
&&J=\frac{n_1+n_2}{2}\:\:\:,\:\:\:J'=\frac{m_1+m_2}{2}\nonumber \\
&&p=\frac{-n_1+n_2}{2}\:\:\:,\:\:\:q=\frac{-m_1+m_2}{2}\:.
\eea
whose limits are $J(0,\infty)$, $J'(0,\infty)$, $p(-\infty,\infty)$, $q(-\infty,\infty)$.
Replacing these variables on Eq.~(\ref{eqn}) one has
\bea
&&e^{-iJ_z \phi_k} \left[ | -i \beta    \rangle_x | i \beta  \rangle_y \otimes
\langle -i \beta  |_x \langle i \beta  |_y \right. \nonumber \\
&&\left. -|  \beta    \rangle_x |  \beta \rangle_y \otimes
\langle  \beta  |_x \langle  \beta  |_y  \right]  e^{iJ_z \phi_k}\nonumber\\
&&=-2 i  e^{-2|\beta|^2}\nonumber \\ &&\times
\sum_{p=-\infty}^{\infty}\sum_{q=-\infty}^{\infty}
\sum_{J=p}^{\infty}\sum_{J'=q}^{\infty}
\nonumber \\
&& \frac{  \beta^{2J} (\beta^*)^{2J'}\:e^{i\frac{\pi}{2}(q-p)} \sin\left[\frac{\pi}{2} (q-p) \right]e^{ i\phi_k(p-q)} }{
 \sqrt{
 \left(J-p \right)!
 \left(J+p \right)!
 \left(J'-q \right)!
 \left(J'+q \right)!
 }
 }\nonumber \\
 &&
|J-p \rangle     |J+p \rangle
\langle J' +q |    \langle J' -q |\:.
\eea
The state $|J-p \rangle     |J+p \rangle$ can be written $|J,p\rangle\rangle$ and \\$|J'-q \rangle     |J'+q \rangle$ written $|J',q\rangle\rangle$.
Now, $\Delta \tilde{\rho}$ is written
\bea
&&\Delta \tilde{\rho}=\frac{1}{M}\sum_{k=0}^{M-1}e^{-i J_z \phi_k} \nonumber \\
&&\left(| -i \beta    \rangle_x | i \beta  \rangle_y \otimes
\langle -i \beta  |_x \langle i \beta  |_y  \right. \nonumber \\
&&-\left. |  \beta    \rangle_x |  \beta \rangle_y \otimes
\langle  \beta  |_x \langle  \beta  |_y \right)
e^{i J_z \phi_k}\:,
\eea
or
\bea
&&\Delta \tilde{\rho}=-2i \frac{e^{-2|\beta|^2}}{M}\sum_{k=0}^{M-1}
\sum_{p=-\infty}^{\infty}\sum_{q=-\infty}^{\infty}
\sum_{J=p}^{\infty}\sum_{J'=q}^{\infty}
\nonumber \\
&& \frac{  \beta^{2J} (\beta^*)^{2J'}\:e^{i\frac{\pi}{2}(q-p)} \sin\left[\frac{\pi}{2} (q-p) \right]
e^{i\phi_k(p-q)} }{
 \sqrt{
 \left(J-p \right)!
 \left(J+p \right)!
 \left(J'-q \right)!
 \left(J'+q \right)!
 }
 }\nonumber \\
 &&
|J,p\rangle\rangle \langle \langle J',q|
\eea
One may group terms in $\Delta \tilde{\rho}$ to express the not-normalized wavefunction
\bea
|\Phi_p\rangle \rangle_u=\sum_J \frac{\beta^{2 J} }{\sqrt{(J-p)!(J+p)!}}|J-p \rangle| J+p\rangle\:,
\eea
and rewrite $\Delta \tilde{\rho}$:
\bea
&&\Delta \tilde{\rho}=-2i \frac{e^{-2|\beta|^2}}{M}\sum_{k=0}^{M-1}
\sum_{p=-\infty}^{\infty}\sum_{q=-\infty}^{\infty} e^{i\frac{\pi}{2}(q-p)} e^{ i \phi_k (p-q)}
\nonumber \\
&&
 \times \sin\left[  \frac{\pi}{2} (q-p) \right]
|\Phi_p\rangle\rangle_u    \langle \langle \Phi_q|_u\:.
\eea
\subsection{Normalization}
In order to normalize $|\Phi_p\rangle \rangle_u$, one has to calculate the normalization factor $\mathcal{N}$:
\bea
|\Phi_p(\phi_k)\rangle \rangle=\mathcal{N}\sum_J \frac{\beta^{2 J}}{\sqrt{(J-p)!(J+p)!}}|J-p \rangle| J+p\rangle.
\eea
Calculate
\bea
 && \langle \langle  \Phi_p |\Phi_p\rangle \rangle=|\mathcal{N}|^2\sum_J \sum_{J'}\nonumber \\&&
 \frac{\beta^{2J}(\beta^*)^{2 J'}}{\sqrt{(J-p)!(J+p)!(J'-p)!(J'-q)!}}\langle \langle |  J',p|J,p\rangle \rangle\:.
\eea
The orthogonality of the number states give
\bea
\langle \langle |  J',p|J,p\rangle \rangle=\delta_{JJ'}\:,
\eea
and therefore
\bea
&& \langle \langle  \Phi_p |\Phi_p\rangle \rangle=|\mathcal{N}|^2\sum_{J=p}^{\infty} \frac{|\beta|^{4J}}{(J-p)!(J+p)!}\:.
\eea
Writing $J-p=Q$ one has
\bea
\label{Q}
&& \langle \langle  \Phi_p |\Phi_p\rangle \rangle=|\mathcal{N}|^2 |\beta|^{4p}\sum_{Q=0}^{\infty} \frac{|\beta|^{4Q}}{Q!(Q+2p)!}=1\:.\nonumber
\eea
One has
\bea
\sum_{Q=0}^{\infty} \frac{|\beta|^{4Q}}{Q!(Q+2p)!}=\frac{1}{(|\beta|^2)^{2 p}}I_{2 p}\left( 2|\beta|^2 \right)\:.
\eea

Calculating $|\mathcal{N}|$, one obtains the normalized function
\bea
|\Phi_p \rangle \rangle=\frac{1}{\sqrt{I_{2p}\left(2 |\beta|^2  \right)}}\sum_{J=p}^{\infty}\frac{\beta^{2J}}{\sqrt{\left(J-p \right)!\left(J+p \right)!}} |J-p\rangle |J+p\rangle.
\eea
A few simplifications lead to
\bea
\label{DeltaFinal2}
\Delta \tilde{\rho}\!\!&=\!\!&\!\!\sum_{p=-\infty}^{\infty}
\sum_{q=-\infty}^{\infty} 2
 e^{- 2|\beta|^2}\left[\frac{1-(-1)^{q-p}}{2}  \right] \sqrt{I_{2p}\left(2 |\beta|^2  \right) I_{2q}\left(2 |\beta|^2  \right)}  \times S(p,q) \times \nonumber \\
&& |\Phi_p\rangle \rangle \langle \langle\Phi_p |\:,
\eea
where
\bea
\label{Spq2}
S(p,q)\equiv \sum_{k=0}^{M-1}e^{-i \phi_k (q-p)}\:.
\eea

{}

\end{document}